\documentclass[12pt]{iopart}
\usepackage{epsfig}

\def\beq{\begin{equation}}
\def\eeq{\end{equation}}
\def\bea{\begin{eqnarray}}
\def\eea{\end{eqnarray}}

\def\ms {\overline{\rm MS}}

\newcommand{\ra}{\rightarrow}
\newcommand{\gsim}{\raisebox{-0.05cm}{$\:\stackrel{>}{{\scriptstyle
 \sim}}\: $} }
\renewcommand{\lsim}{\raisebox{-0.05cm}{$\:\stackrel{<}{{\scriptstyle
 \sim}}\: $} }
\newcommand\MSb{$\overline{\mbox{MS}}$}

\begin{document}
\addtolength{\topmargin}{-48pt}

\title[]
      {Parton densities and structure functions 
      at next-to-next-to-leading order and beyond}

\author{W.L. van Neerven and A. Vogt
}

\address{Instituut-Lorentz, University of Leiden, \\
         P.O. Box 9506, 2300 RA Leiden, The Netherlands}

\begin{abstract}
We summarize recent results on the evolution of unpolarized parton
densities and deep-inelastic structure functions in massless
perturbative QCD.
Due to last year's extension of the integer-moment calculations of the
three-loop splitting functions, the NNLO evolution of the parton
distributions can now be performed reliably at momentum fractions
$x\,\gsim\, 10^{-4}$, facilitating a considerably improved theoretical
accuracy of their extraction from data on deep-inelastic scattering.
The NNLO corrections are not dominated, at relevant values of $x$, by 
their leading small-$x$ terms.
At large~$x$ the splitting-function series converges very rapidly,
hence, employing results on the three-loop coefficient functions, the
structure functions can be analysed at N$^3$LO for $x >\! 10^{-2}$.
The resulting values for $\alpha_s$ do not significantly change beyond
NNLO, their renormalization scale dependence reaches about $\pm 1\%$ at
N$^3$LO.
\end{abstract}



\maketitle

\section{Introduction}

Precise predictions for hard strong-interaction processes require 
transcending the standard next-to-leading order (NLO) approximation of 
perturbative QCD. Resumm\-ations of large logarithms may be sufficient 
for specific processes, but generally full next-to-next-to-leading 
order (NNLO) calculations are called for. For electron--proton 
scattering and proton--(anti-)proton colliders this demands both 
partonic cross sections and parton distribution of NNLO accuracy.

The former quantities are presently available only for the structure 
functions in deep-inelastic scattering (DIS)~\cite{c2DIS} -- which 
provide the backbone of our knowledge of the proton's parton densities 
and are among the quantities best suited for measuring~$\alpha_s$ -- 
and the total cross section for the Drell-Yan process~\cite{c2DY} -- 
which in the form of $W$ and $Z$ production is an excellent candidate 
for an accurate luminosity monitor at {\sc Tevatron} and the LHC~%
\cite{Wprod}. The calculation of other processes like jet production at 
NNLO is under way, e.g., the required two-loop two-to-two matrix 
elements have been computed~\cite{matel} using the pioneering results~%
\cite{dbox} for the scalar double-box diagrams. See~refs.~\cite{NGMrd} 
for recent brief overviews. 
Partial NNLO results (the soft- and virtual-gluon contributions) have 
also been obtained for Higgs production via gluon-gluon fusion in the
heavy top-quark limit~\cite{Hsoft}, see also ref.~\cite{KLS}.

The three-loop splitting functions entering the NNLO evolution of the 
parton distributions have not been completed so far either~\cite{P2ex}.
However, previous partial results~\cite{moms,lowx} have been substanti%
ally extended by the calculation of two more Mellin moments~\cite{RV00}.
In Section 2 we discuss the resulting improvement~\cite{NV3} of our 
approximations of the splitting functions in $x$-space~\cite{NV12}, 
and compare, in the extended range $x\,\gsim\, 10^{-4}$ of safe 
applicability, the resulting approximate NNLO flavour-singlet evolution 
and its scale stability to the NLO results. We also briefly re-address 
\cite{vN93,BNRV} the question to what extent the leading small-$x$ 
contributions to the splitting functions and coefficient functions 
dominate the small-$x$ evolution. 

Taking into account the fast convergence of the splitting-function 
series shown in Section 2, the next-to-next-to-next-to-leading order
(N$^3$LO) corrections for the DIS structure functions can be effectively
derived at $x\!\! >\!\! 10^{-2}$ using available partial results~\cite
{moms,RV00,avsg} for the three-loop coefficient functions. 
The effect of the NNLO and N$^3$LO terms is discussed in Section~3 
for the scaling violations of the non-singlet structure function $F_2$ 
and the resulting determination of $\alpha_s$~\cite{NV4}. Here we also 
illustrate the predictions of the principle of minimal sensitivity 
\cite{PMS}, the effective charge method~\cite{ECH} and the Pad\'e 
summation~\cite{Pade} which in this case, unlike the soft-gluon 
resummation, seem to facilitate a reliable estimate of the corrections 
even beyond N$^3$LO.

A first study has been performed~\cite{MRST} of the effects of the NNLO 
corrections (using our original approximations~\cite{NV12} for the 
three-loop splitting functions) in a global parton analysis. 
See refs.~\cite{KKPS,SaYn} for beyond-NLO $\alpha_s$ analyses of DIS 
data using methods more directly based on the integer-moment results
\cite{moms,RV00}.

\section{Singlet parton densities and structure functions at NNLO}

We first illustrate our approximation procedure for the three-loop  
splitting functions $P^{(2)}$. As an example we discuss the $N_f^{\,1}$ 
term $P_{qg,1}^{(2)}$ of the gluon-quark splitting function $P_{qg}$ 
dominating the small-$x$ evolution of the quark densities. 
In the \MSb\ scheme employed in our studies, this function can be 
written as
\beq
\fl
\label{eq1} \quad\quad\quad
  P_{qg,1}^{(2)}(x) \, =\, \sum_{m=1}^{4}\! A_m \ln^m (1-x)
  + f_{\rm smooth}(x) + \sum_{n=1}^{4} B_n \ln^n x 
  + \sum_{p=0}^{1}  C_p\, \frac{\ln^p x}{x} \:\: .
\eeq
The leading small-$x$ coefficient $C_1$ has been derived by Catani 
and Hautman~\cite{lowx}. The function $f_{\rm smooth}$ collects all 
contributions which are finite for $0 \leq  x \leq 1$. 
This regular term constitutes the mathematically complicated part of 
Eq.~(\ref{eq1}), involving higher transcendental functions like the 
harmonic polylogarithms~\cite{hpol}. 

For our improved approximations~\cite{NV3} we choose three or two of 
the large-$x$ logarithms in Eq.~(\ref{eq1}), a one- or two-parameter 
smooth function (low powers or simple polynomials of $x$) and two of 
the small-$x$ terms ($x^{-1}$ together with $\ln x$ or $\ln^2 x$). 
Their coefficients are then determined from the six even-integer 
Mellin moments
\beq
\label{eq1b} 
  P_{qg,1}^{(2)}(N) \, =\, \int_0^1 \! dx\, x^{N-1}\, P_{qg,1}^{(2)}(x) 
\eeq
computed by 
Larin et al.~\cite{moms} and Retey and Vermaseren~\cite{RV00}. By 
varying these choices we arrive at about 50 approximations (see Fig.~1
of ref.~\cite{NV3}). The two functions spanning the resulting error 
band for most of the $x$-range are finally selected as our best 
estimates for $P_{qg,1}^{(2)}(x)$ and its residual uncertainty. 

These two functions, denoted by `A' and `B', are shown in Fig.~1 
together with their (practically indistinguishable) real moments 
(\ref{eq1b}) for $2 \,\lsim\, N \leq 30$ and their convolutions 
\beq
\label{eq1c}
  [ P_{qg}^{(2)} \otimes g ](x) \, = \, \int_x^1 \! \frac{dy}{y} 
  \, P_{qg}^{(2)}(y)\, g\bigg(\frac{x}{y}\bigg) 
\eeq
with a typical gluon distribution $g$. In Figs.~1(b) and 1(c) the 
corresponding results for the $N_f^{\, 2}$ term have been supplemented 
for $N_f\! =\! 4$. Note that, like refs.~\cite{moms,RV00}, we use the 
small expansion parameter $a_s\! =\! \alpha_s/(4 \pi)$; scaling down 
the ordinates by a factor 2000 yields the results for an expansion in 
$\alpha_s$.  The large impact of the $N\! =\! 10$ and 12 moments
\cite{RV00} is illustrated by also showing our less accurate, but 
compatible original approximations~\cite{NV12} based on the four lowest 
even-integer moments~\cite{moms}.

Knowing the leading $x^{-1} \ln x$ term~\cite{lowx} is clearly 
instrumental in constraining the small-$x$ behaviour of $P_{qg,1}^{(2)}$
-- something not efficiently done by a small number of $N\geq 2$ integer
moments (\ref{eq1b}). However, even at $x \leq 10^{-3}$ where the 
non-$x^{-1}$ parts contribute less than 10\% to both approximations `A' 
and `B', this term does not sufficiently dominate over the (so far 
uncalculated) subleading $C_0\, x^{-1}$ contribution in Eq.~(\ref{eq1}),
leaving us with a sizeable uncertainty of $P_{qg,1}^{(2)}$ for 
$x\,\lsim\, 10^{-2}$. We will examine the dominance of the $x^{-1} 
\ln x$ and $x^{-1}$ terms for the convolution (\ref{eq1c}), which in 
any case considerably smoothes out the oscillating differences of the 
approximations, at the end of this section. 

\begin{figure}[tbp]
\vspace*{1mm}
\centerline{\hspace*{-1mm}\epsfig{file=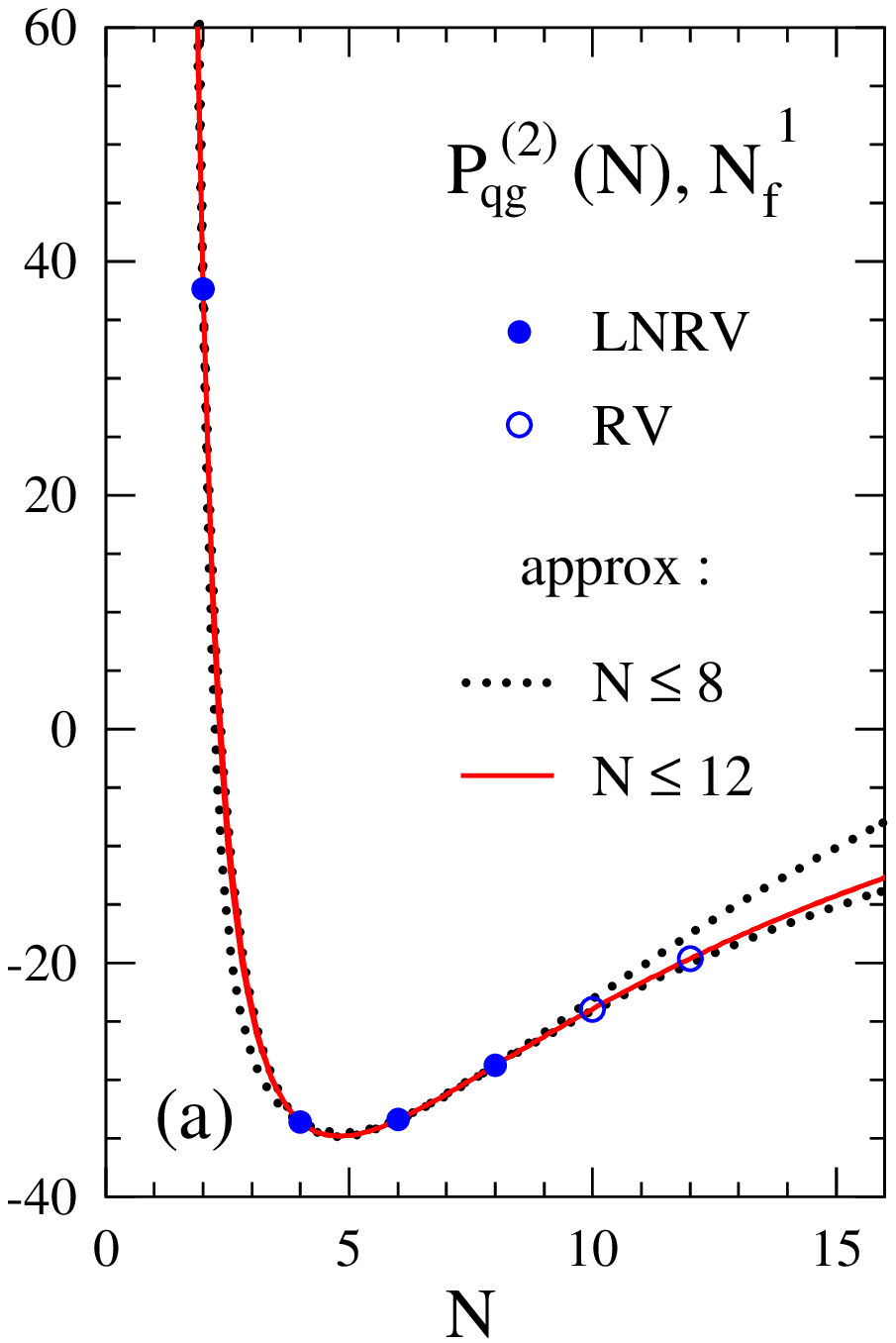,width=5.2cm,angle=0}
\epsfig{file=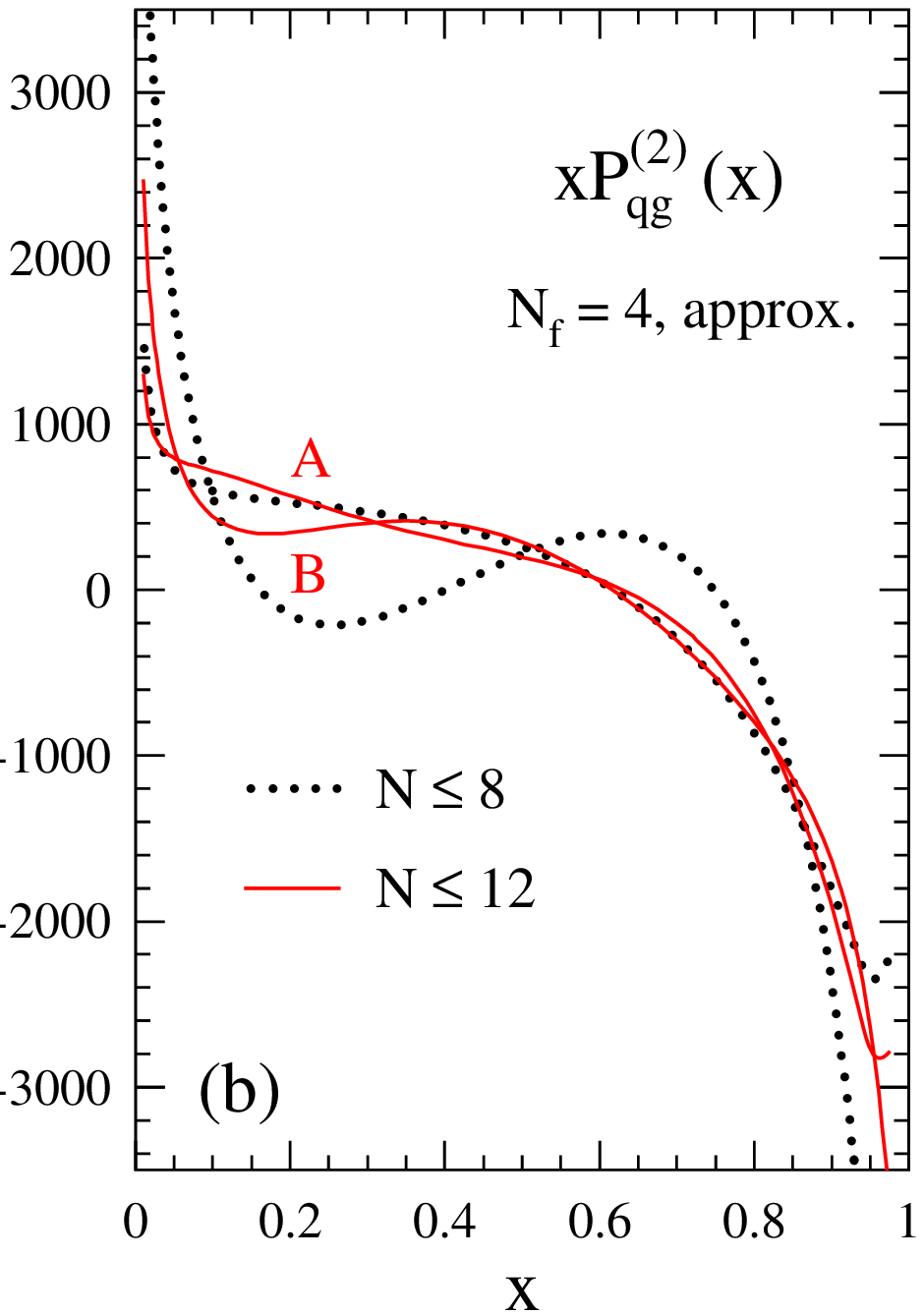,width=5.2cm,angle=0}    
\epsfig{file=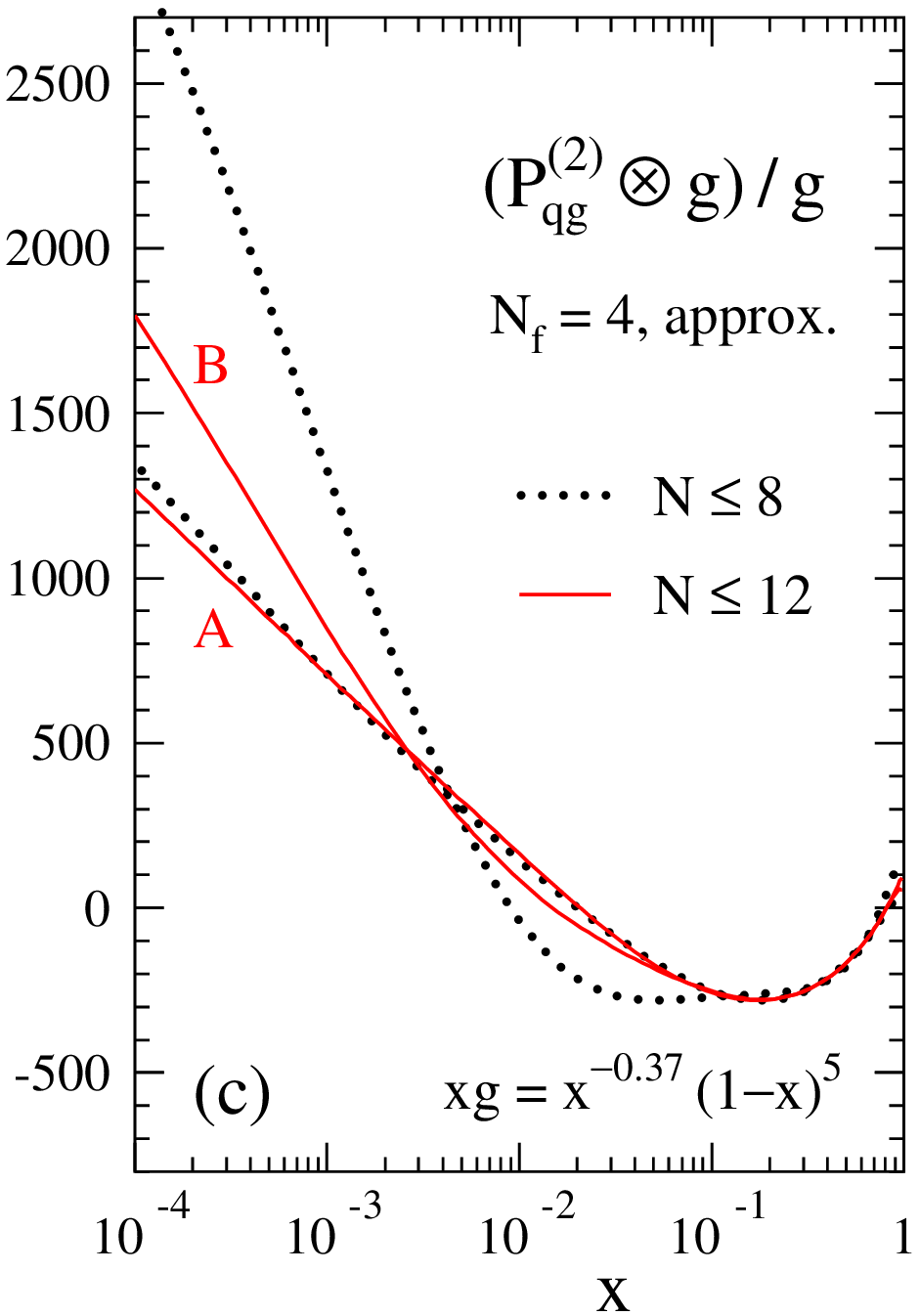,width=5.2cm,angle=0}}
\caption{
 {\bf (a)} Exact results~\protect\cite{moms,RV00} (points) and 
 approximations~\protect\cite{NV12,NV3} (curves) for the moments of 
 the $N_f^1$ term of the three-loop gluon-quark splitting function
 in the \MSb\ scheme. 
 {\bf (b)} The approximations in $x$-space for $N_f\! =\! 4$, and 
 {\bf (c)} their normalized convolutions with a typical gluon 
 distribution.
 }
\end{figure}

After applying analogous procedures to the other three-loop splitting
functions (see Figs.~2 and 3 of ref.~\cite{NV3}) we are ready to 
exemplify the effect of the NNLO contributions on the evolution of the 
parton densities. This is done in Fig.~2 via the logarithmic 
factorization-scale derivatives $\dot{f} = d\ln f / d\ln \mu_f^2$ of
the typical singlet quark ($\Sigma$) and gluon ($g$) distributions
\bea
\label{eq2}
  x\Sigma (x,\mu_{f,0}^2) &=& 0.6\, x^{-0.3}\,\, (1-x)^{3.5}
                                        (1 + 5\, x^{0.8})
  \nonumber \\
  xg (x,\mu_{f,0}^2) &=& 1.0\, x^{-0.37} (1-x)^{5} 
\eea
at the reference scale
\beq
\label{eq3}
  \mu_f^2 \, =\, \mu_{f,0}^2 \,\approx\, 30 \mbox{ GeV}^2
  \quad \longleftrightarrow \quad
  \alpha_s (\mu_{f,0}^2)\, =\, 0.2 \:\: .
\eeq
In Fig.~2(a) the resulting relative NNLO effects are shown for the 
standard choice $\mu_r = \mu_f$ of the renormalization scale. Note that 
the spikes close to $x = 0.1$ do not represent large NNLO corrections, 
instead they derive from zeros of the denominators. In fact, the NNLO 
corrections are very small (except for the quark evolution at small~%
$x$), much smaller than their NLO counterparts not shown in the figure 
(except for the gluon evolution at small $x$, where already the NLO 
corrections are rather small). Consequently, as illustrated in Figs.\ 
2(b) and 2(c), the renormalization-scale stability of the prediction is
considerably improved over the full $x$-range. If measured over the 
conventional range $1/2\,\mu_f\!\leq\!\mu_r\!\leq 2\,\mu_f$, the $\mu_r$
dependence of $\dot{\Sigma}$ amounts to less than $\pm 2\%$ at large 
$x$ and $\pm 5\%$ small $x$, that of $\dot{g}$ to less than $\pm 1\%$
and $\pm 2\%$, respectively.
  
\begin{figure}[tbp]
\vspace*{1mm}
\centerline{\hspace*{-1mm}\epsfig{file=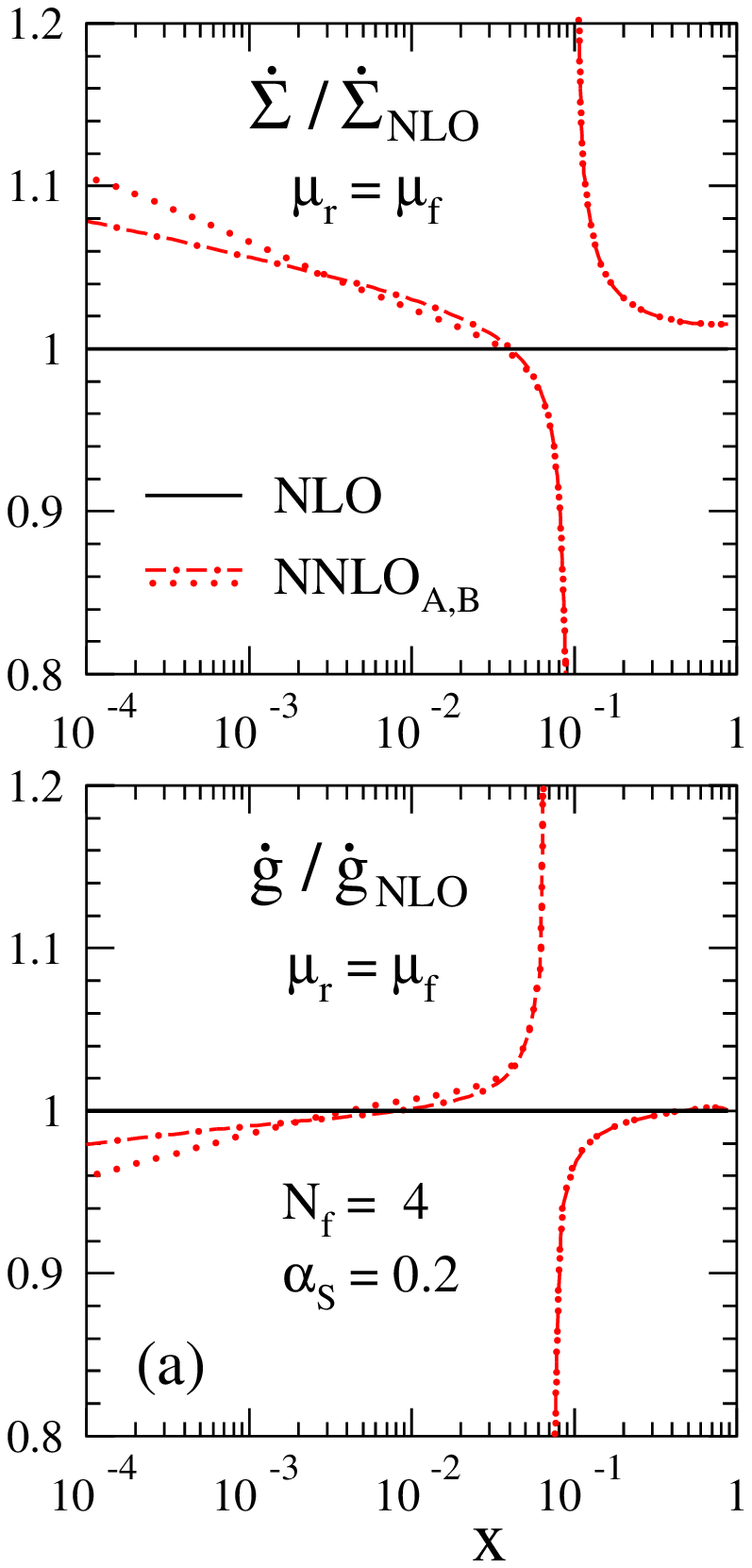,width=5.2cm,angle=0}
\epsfig{file=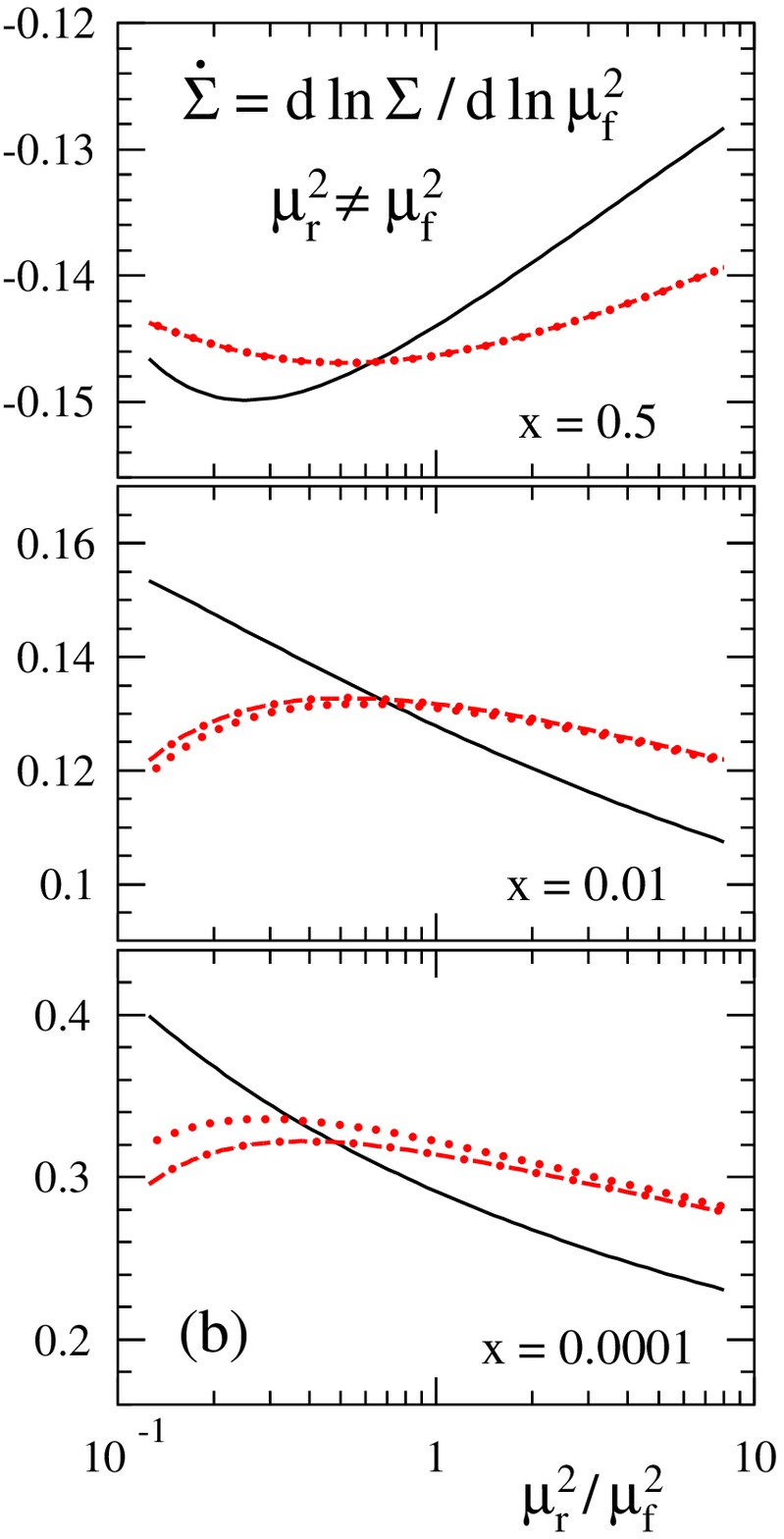,width=5.2cm,angle=0}
\epsfig{file=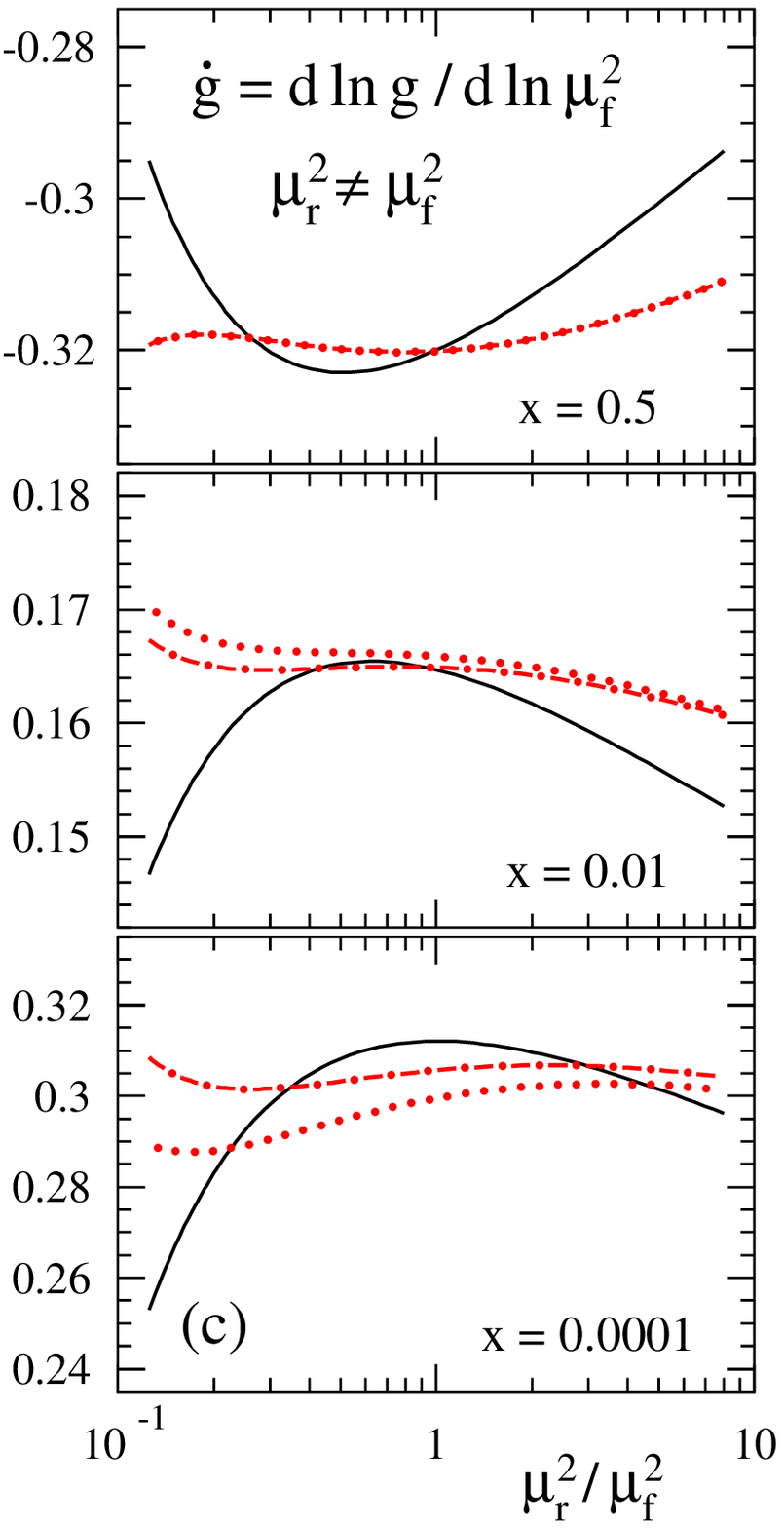,width=5.2cm,angle=0}}
\caption{
 {\bf (a)} Size and present approximation uncertainties of the NNLO 
 corrections to the scale derivatives of the singlet quark and gluon 
 densities for the input (\protect\ref{eq2}) and 
 (\protect\ref{eq3}) at $\mu_r\! =\! \mu_f$. 
 {\bf (b,c)} The renormalization scale dependence at NLO and NNLO for
 three typical values~of~$x$.  
 }
\label{avf2}
\end{figure}

Also displayed in Fig.~2 are the present inaccuracies (`A'~vs.~`B') of 
the NNLO results caused by the remaining uncertainties~\cite{NV3} of 
the three-loop splitting functions. These inaccuracies are entirely 
negligible at large $x$. Down to $x \simeq 10^{-4}$ they amount to 
about $\pm 2\%$ or less with respect to the central results not shown 
in the figure, even if the bands in Fig.~2(a) are increased by 50\% in 
order to account for any possible underestimate of the uncertainties. 

At lower scales the splitting-function uncertainties have a larger
impact, mainly due to the larger value of $\alpha_s$.  For example, the 
spread corresponding to Fig.~2(a) reaches about $\pm 4\%$ for 
$\dot{\Sigma}$ and $\pm 3\%$ for $\dot{g}$ at $x\! =\! 10^{-4}$ and 
$\mu_{f}^2\approx 3 \mbox{ GeV}^2$ corresponding to $\alpha_s = 0.3$. 
In view of the also enhanced NLO scale dependence, the approximate 
NNLO evolution represents an improvement over the NLO treatment even 
with inaccuracies of this size.

The electromagnetic singlet structure function $F_2$ and its $Q^2$
derivatives are presented in Fig.~3 for the parton densities (\ref{eq2})
at $Q^2 = \mu_{f,0}^2 \approx 30 \mbox{ GeV}^2$. The large NNLO 
corrections at very large $x$ originate in the non-singlet part of the 
two-loop quark coefficient function. 
Note, however, that the (positive) gluon contribution to $dF_{2,S} / 
d\ln Q^2$ at NNLO still amounts to 5\% at $x\! =\! 0.5$ (40\% more than 
at NLO), and falls below 1\% only above $x \! =\! 0.7$~\cite{NV12}. 
This effect is large enough to jeopardize analyses applying a 
non-singlet formalism to the proton's $F_2$ in the region $x > 0.3$.

\begin{figure}[tbp]
\vspace*{1mm}
\centerline{\hspace*{-1mm}\epsfig{file=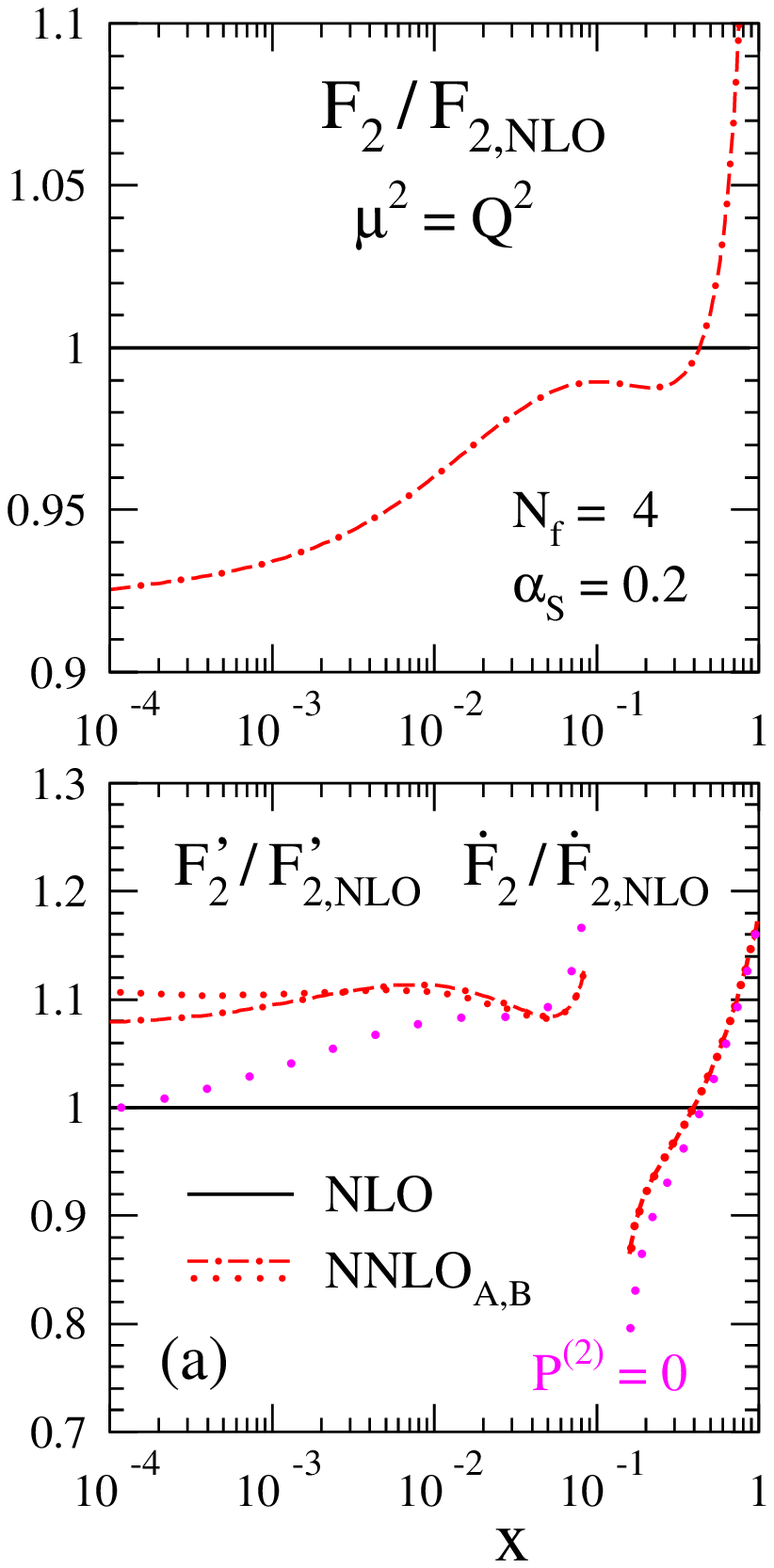,width=5.2cm,angle=0}
\epsfig{file=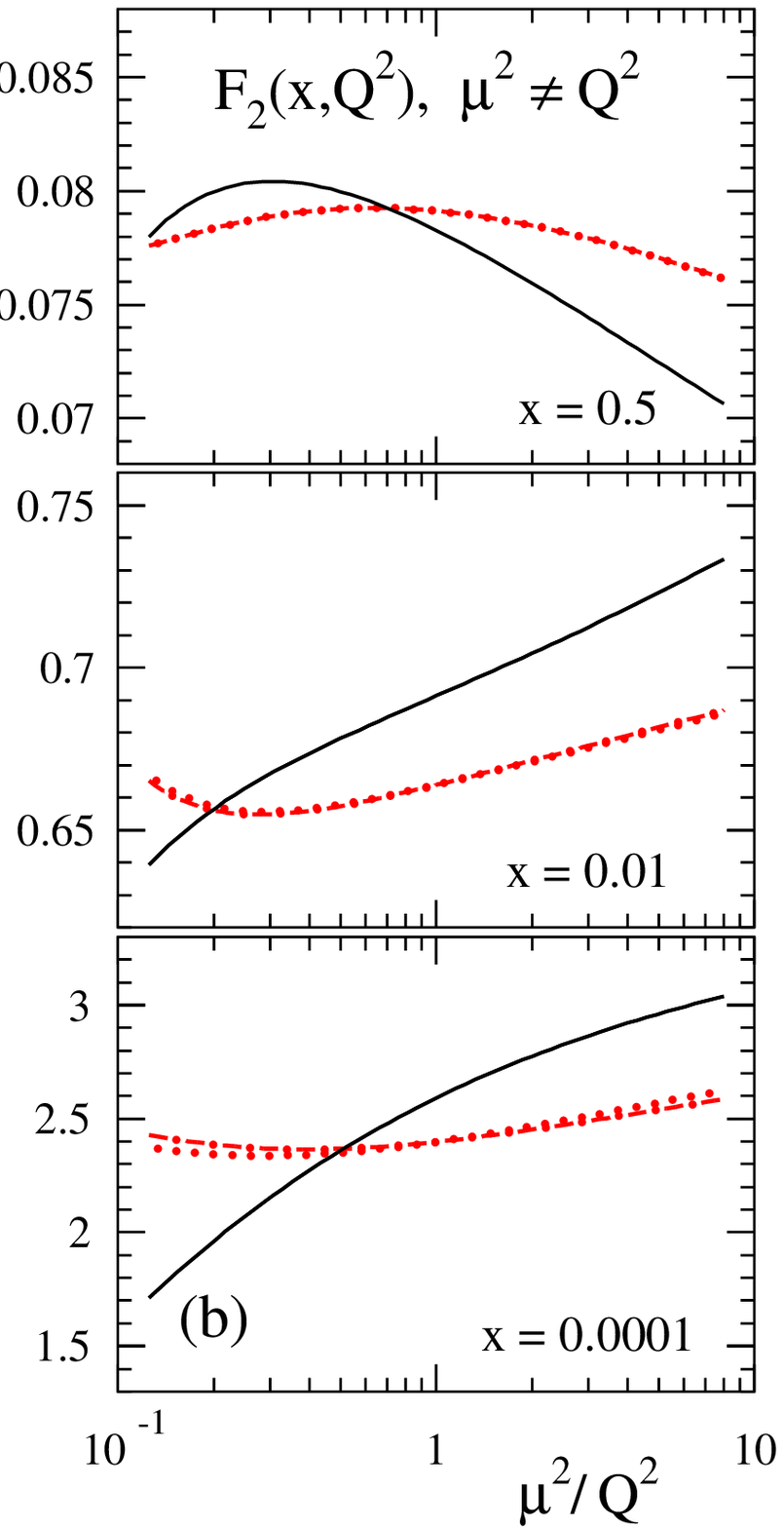,width=5.2cm,angle=0}
\epsfig{file=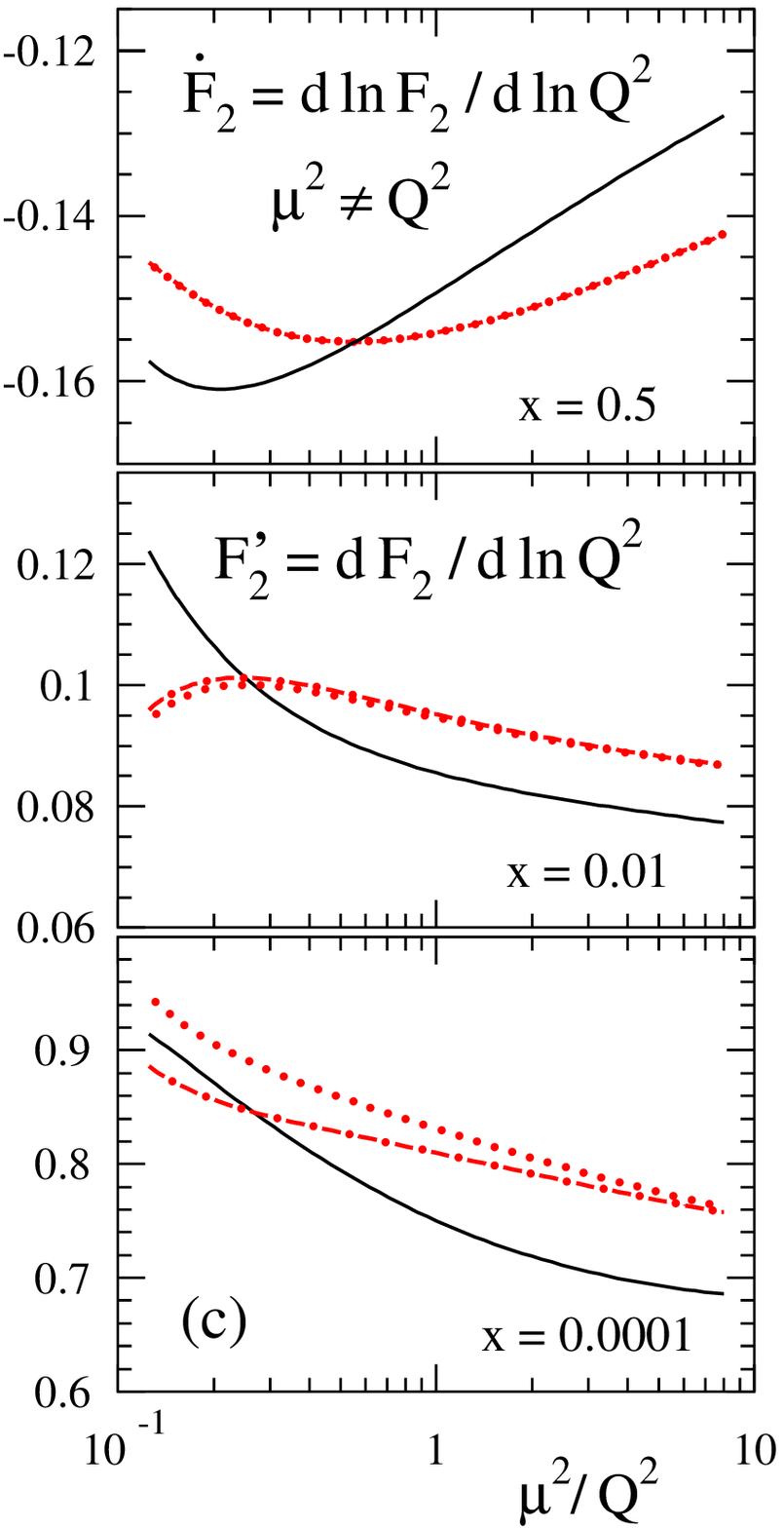,width=5.2cm,angle=0}}
\caption{
 {\bf (a)} The NNLO corrections for the singlet $F_2$ and its $Q^2$ 
 derivatives (linear at small $x$, logarithmic at large $x$) for the 
 input (\ref{eq2}) at $\mu_r^2 = \mu_f^2 \equiv \mu^2 = \mu_{f,0}^2 = 
 Q^2 \approx 30 \mbox{ GeV}^2$. 
 {\bf (b,c)} The scale dependence at NLO and NNLO for three typical
 values of~$x$.
 }
\label{avf3}
\end{figure}

The negative NNLO corrections to $F_2$ at small $x$ arise from the 
two-loop gluon coefficient function $c_{\, 2,g}^{\, (2)}$. The $Q^2$ 
derivative, on the other hand, receives a +10\% NNLO correction for 
$10^{-4}\,\lsim\, x \,\lsim\, 10^{-2}\,$; its break-up is illustrated 
in Fig.~3(a) by the results for $P^{(2)} = 0$.
Also for $F_2$ and its scaling violations the inclusion of the NNLO
terms leads to a substantial decrease of the scale uncertainties as 
shown in Figs.~3(b,c), which facilitates more precise extractions of 
the parton distributions from data on these quantities.

\begin{figure}[tbp]
\vspace*{1mm}
\centerline{\hspace*{-1mm}\epsfig{file=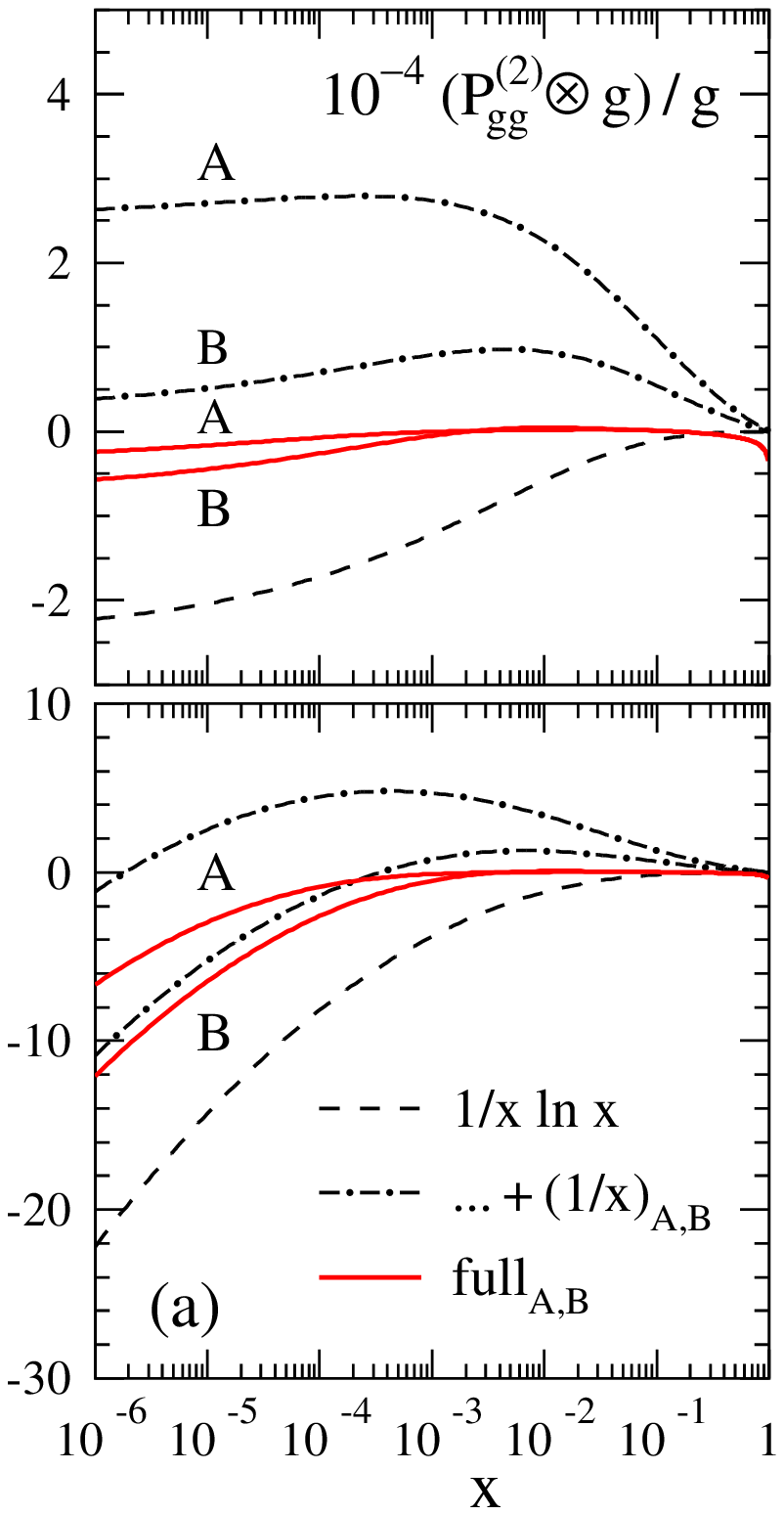,width=5.2cm,angle=0}
\epsfig{file=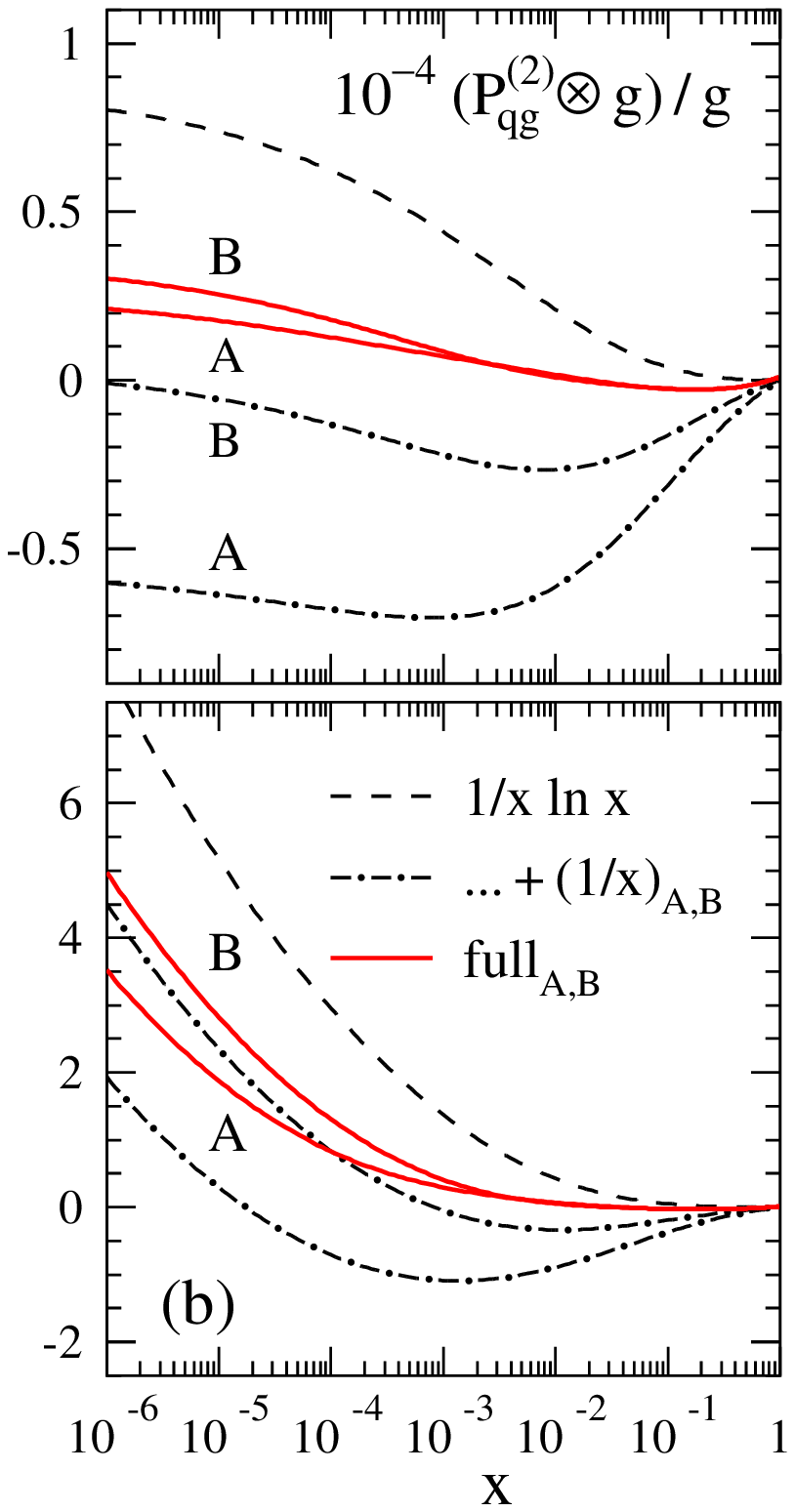,width=5.2cm,angle=0}
\epsfig{file=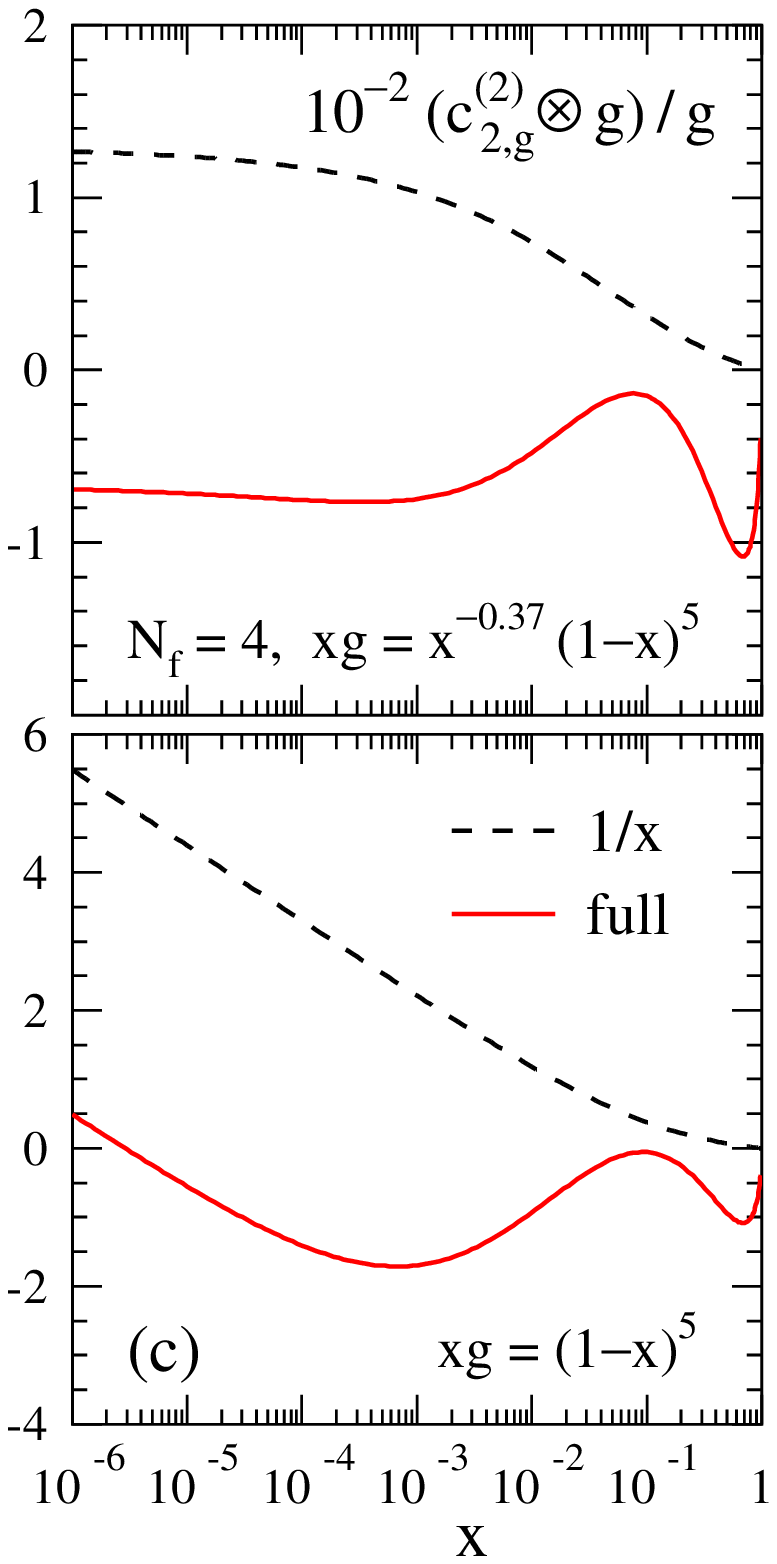,width=5.2cm,angle=0}}
\caption{
 The convolutions of the leading (and, for $P^{(2)}$, subleading) 
 small-$x$ terms of {\bf (a,b)} the NNLO splitting functions 
 $P_{gg}^{(2)}$ and $P_{qg}^{(2)}$ and {\bf (c)} the NNLO coefficient
 function $c_{\, 2,g}^{\, (2)}$ with two typical `steep' (upper row)
 and `flat' (lower row) gluon distributions, compared to the respective
 (for $P^{(2)}$ approximate) full results.
 }
\label{avf3a}
\end{figure}

We conclude this section by examining the dominance of the small-$x$ 
terms of the NNLO splitting functions and coefficient functions for the 
small-$x$ convolutions. In Fig.~4 the results for $ f \otimes g$, 
$f = P_{gg}^{(2)}$, $P_{qg}^{(2)}$ and $c_{\, 2,g}^{\, (2)}$ obtained 
by keeping only the $x^{-1} \ln^k x$~terms of $f$ are compared, down 
to $x = 10^{-6}$, with the (for $P^{(2)}$ approximate) full results. 
The dependence on the gluon distribution $g$ is illustrated by 
employing, besides our `steep' standard input (\ref{eq2}), also a 
low-scale `flat' shape, $xg \sim x^0\,$ for $\,x \ra 0$. 

Keeping only the leading $x^{-1} \ln x$ terms~\cite{lowx} of 
$P_{gg}^{(2)}$ and $P_{qg}^{(2)}$ does not lead to reasonable 
approximations as shown in Figs.~4(a,b), regardless of the gluon 
distribution. Even for the more favourable flat shape the offsets 
amount to about a a factor of two even at $x = 10^{-6}$. Besides the 
$x^{-1}$ contributions, the \mbox{non-$x^{-1}$} terms do not seem to be 
sufficiently suppressed either, at least for a steep gluon distribution.
Due to the present large uncertainties~\cite{NV3} on the $x^{-1}$ terms,
however, definite conclusions especially for a flat gluon distribution 
require the computation~\cite{P2ex} of the exact $x$-dependence of 
$P_{ij}^{(2)}$.
On the other hand, such conclusions can be drawn already~\cite{vN93} 
for the convolutions of the two-loop coefficient function $c_{\,2,g}
^{\,(2)}$~\cite{c2DIS} shown in Fig.~4(c). The leading $x^{-1}$ term
\cite{lowx} does not dominate over the non-$x^{-1}$ contributions
at any $x$-values of practical interest. 

\section{Non-singlet structure functions at NNLO and beyond}

The scaling violations of the non-singlet structure functions 
$F_{a,\rm NS}$, $a = 1,2,3$, can be conveniently discussed in terms of 
the physical evolution kernels $K_{a,\rm NS\,}$, 
\beq
\label{eq4}
  \frac{d}{d \ln Q^2} \, F_{a,\rm NS} \, = \,
  K_{a,\rm NS} \otimes F_{a,\rm NS}   \, = \,
  \sum_{l=0} a_s^{\, l+1}\, K_{a,l} \otimes F_{a,\rm NS} \:\: .
\eeq
The N$^l$LO expansion coefficients $K_{a,l}$ are combinations of the
coefficient functions up to $l$ loops and the splitting functions up to
$l\! +\! 1$ loops. See Eqs.~(2.7) -- (2.9) of ref.~\cite{NV4} for the 
details. The advantage of Eq.~(\ref{eq4}) is that any dependence on the
factorization scheme and the scale $\mu_f$ has been eliminated 
explicitly. Note, however, that in $\alpha_s$ analyses in terms of 
coefficient functions and parton distributions the main uncertainty
arises from the dependence on the renormalization scale $\mu_r$, not the 
factorization scale $\mu_f$.

\begin{figure}[tb]
\vspace*{1mm}
\centerline{\hspace*{-1mm}\epsfig{file=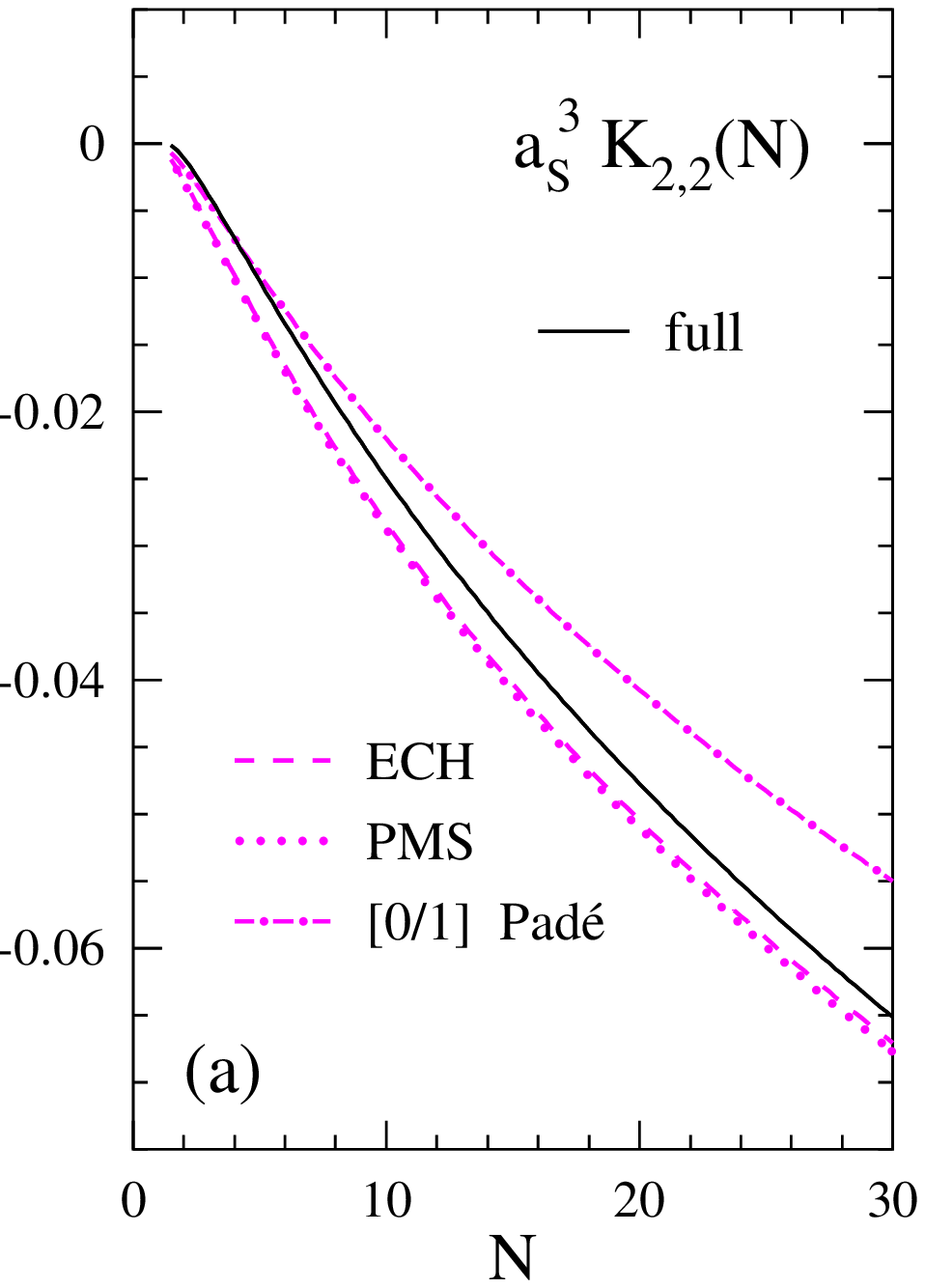,width=5.2cm,angle=0}
\epsfig{file=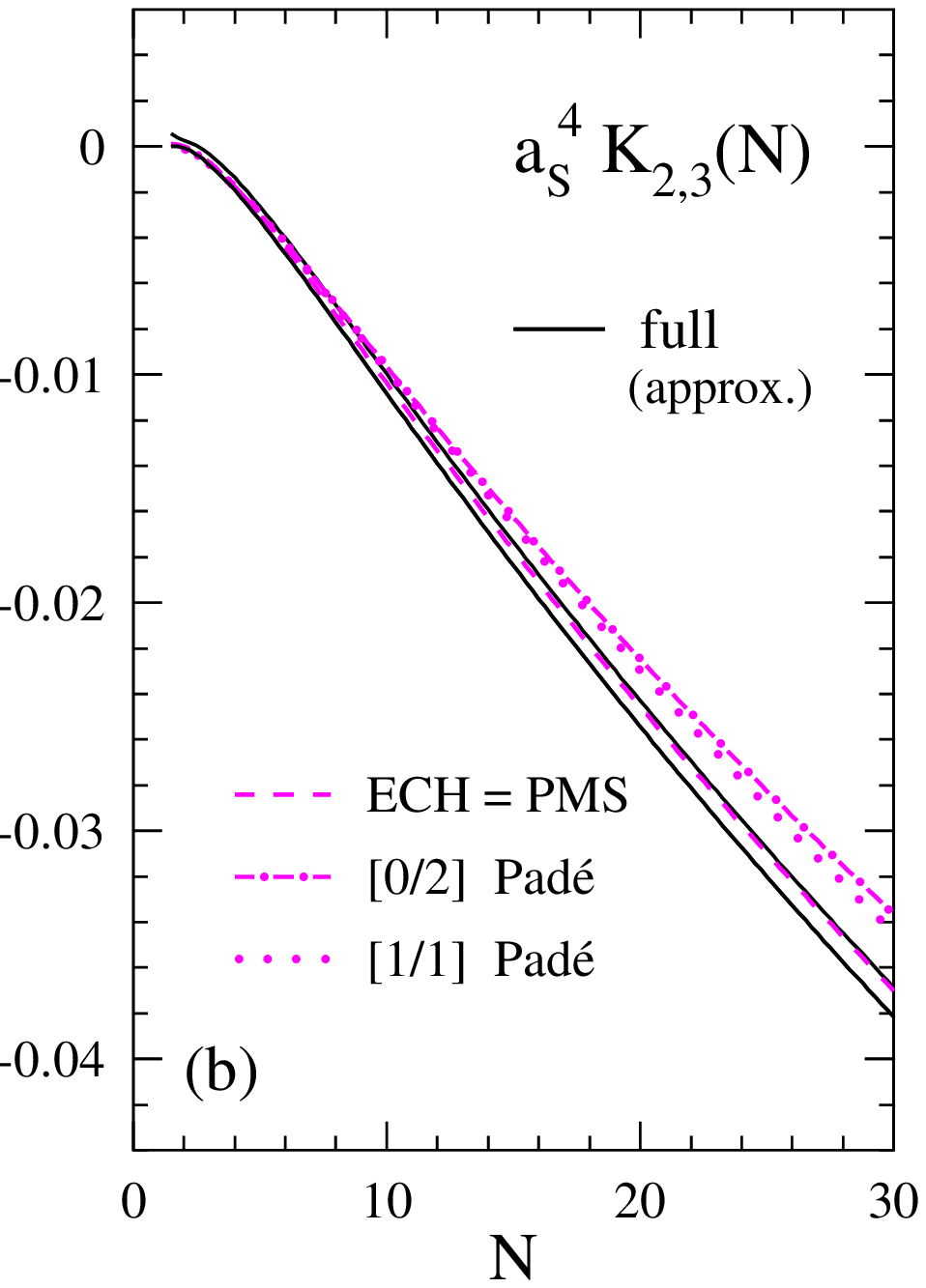,width=5.2cm,angle=0}
\epsfig{file=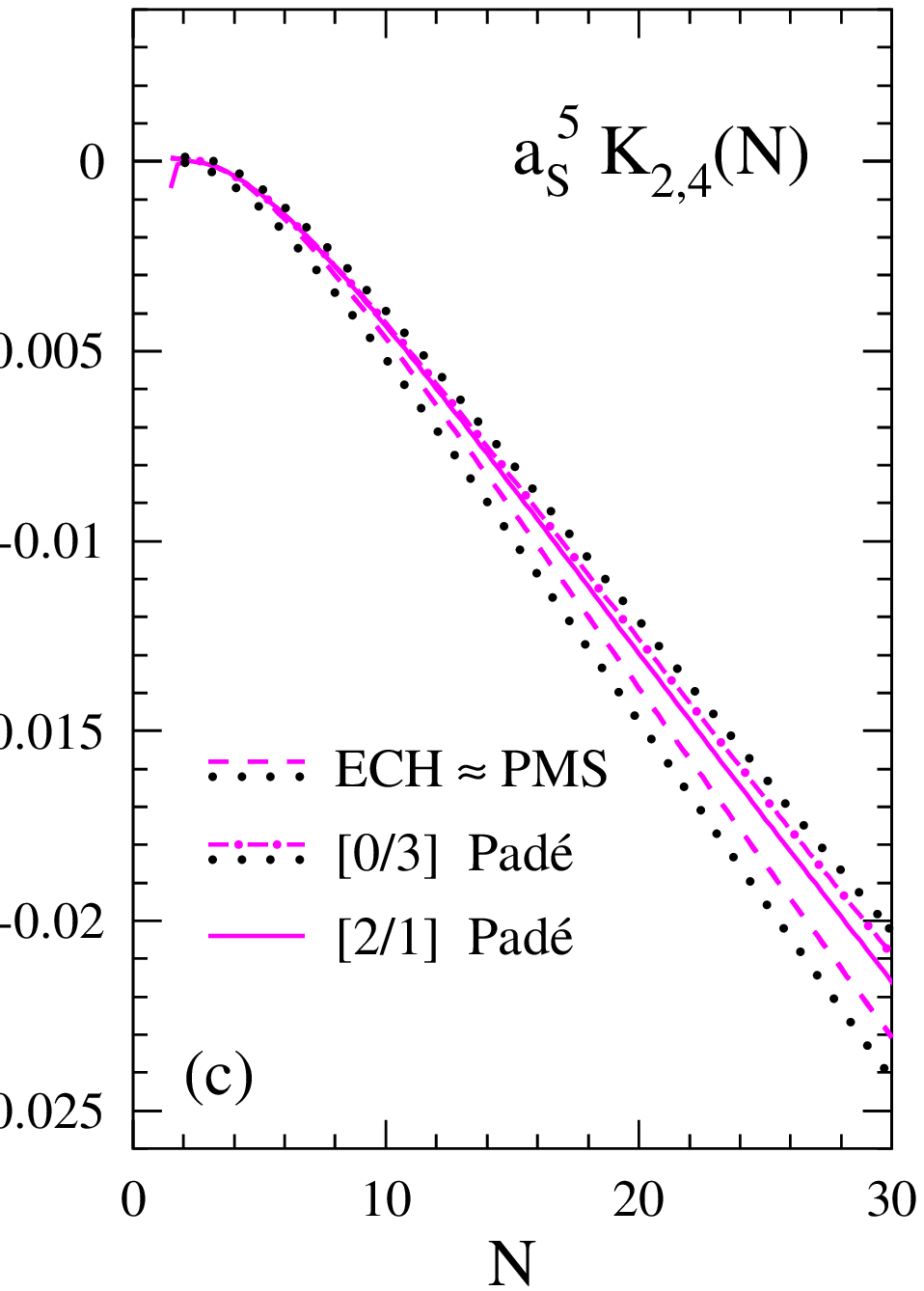,width=5.2cm,angle=0}}
\caption{The PMS, ECH and Pad\'e estimates of {\bf (a)} the NNLO, 
 {\bf (b)} the N$^3$LO, and  {\bf (c)} the N$^4$LO contributions to 
 the $N$-space evolution kernel for $\frac{1}{x} F_{2,\rm NS}$ at 
 $\mu_r^2 = Q^2$, $N_f = 4$ and $\alpha_s = 0.2$. Also shown are the 
 (approximate) full NNLO and N$^3$LO results.}
\label{avf5}
\end{figure}

The NNLO kernels $K_{a,2}$ are fixed by the two-loop coefficient 
functions~\cite{c2DIS} (for which compact parametrizations are available
\cite{NV12}) together with our approximations~\cite{NV3} for three-loop 
splitting functions. The uncertainties of the latter are negligible in 
the region $x >\! 10^{-2}$ in which we are mainly interested for the 
non-singlet case. 
 
Due to the fast large-$x$ convergence of the splitting function series 
illustrated in Fig.~2, the N$^3$LO kernels are dominated by the 
coefficient functions. The seven lowest even (for $F_{1,2}$) and odd 
(for $F_{3}$) moments at three loops have been computed by Larin et~al.\
\cite{moms} ($N \leq 10$ for $F_{1,2}$) and Retey and Vermaseren~%
\cite{RV00}. When complemented by the four leading large-$x$ terms 
$\, (1-x)^{-1} \ln^k (1-x)$, $k = 2,\,\ldots ,\, 5\, $ derived~\cite
{avsg} from the soft-gluon resummation \cite{sglue}, this information 
facilitates $x$-space approximations (analogous to those for $P^{(2)}$ 
exemplified in Section~2) which are sufficiently accurate in the 
above-mentioned region of $x$, as shown in Figs.~2 and 3 of 
ref.~\cite{NV4}. 
The small effect of the uncalculated four-loop splitting functions 
$P^{(3)}$ can be estimated by assigning a 100\% error to the Pad\'e 
prediction $P_{\rm NS}^{(3)}(N)\approx P_{\rm NS}^{(3)}(N){}_{\, [1/1]
\:\rm Pad\acute{e}\,}$. Actually the small residual uncertainties of 
$K_{a,3}$ at $x >\! 10^{-2}$ are dominated by this error, not by the
approximation uncertainties~\cite{NV4} of the three-loop coefficient 
functions.

Knowing the evolution kernels to such a high order also facilitates a
test of the predictions for $K_{a,l}$ derived from the Pad\'e summation
\cite{Pade} of the perturbation series and from renormalization-scheme 
optimizations like the principle of minimal sensitivity (PMS)~\cite{PMS}
and the effective charge (ECH) method~\cite{ECH}. In Figs.~5(a) and 
5(b) these predictions are compared to the (approximate) full kernels
$K_{2,2}$ and $K_{2,3}$ for $F_{2,\rm NS}$ in $N$-space. The estimates 
by all these methods agree rather well for the $\alpha_s^5$ (N$^4$LO) 
contribution shown in Fig.~5(c). Such an agreement is usually 
interpreted~\cite{Pade} as evidence for the approximate correctness of 
the predictions. 

The large-$N\, /\,$large-$x$ behaviour of the kernels $K_{a,\rm NS}$ is 
dominated by the soft-gluon terms $a_s^{l+1} \ln^k N$, $1 \leq k \leq 
l\! +\! 1$. The soft-gluon resummation~\cite{sglue} at next-to-next-%
to-leading logarithmic accuracy~\cite{avsg} fixes the coefficients of 
the leading three terms at all orders. However, as shown in Figs.~8 and 
9 of ref.~\cite{NV4}, these contributions do not provide quantitative 
predictions for $K_{a,l}$ at practically relevant values of $N\, /\, x$,
due to the (quite generally, see ref.~\cite{GR01}) large coefficients 
of subleading logarithms at higher orders~\cite{avsg}. 
 
\begin{figure}[tb]
\vspace*{1mm}
\centerline{\hspace*{-1mm}\epsfig{file=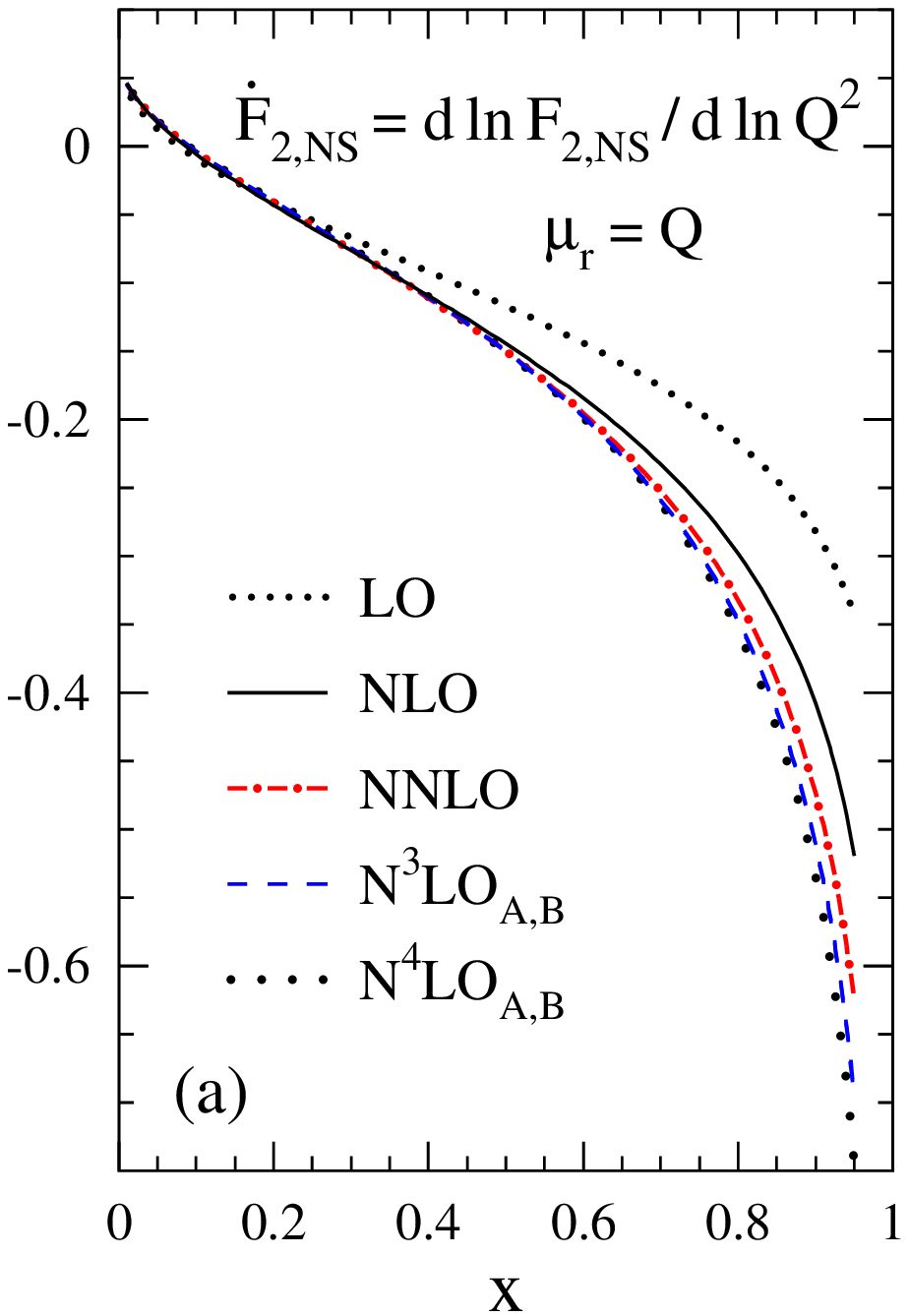,width=5.2cm,angle=0}
\epsfig{file=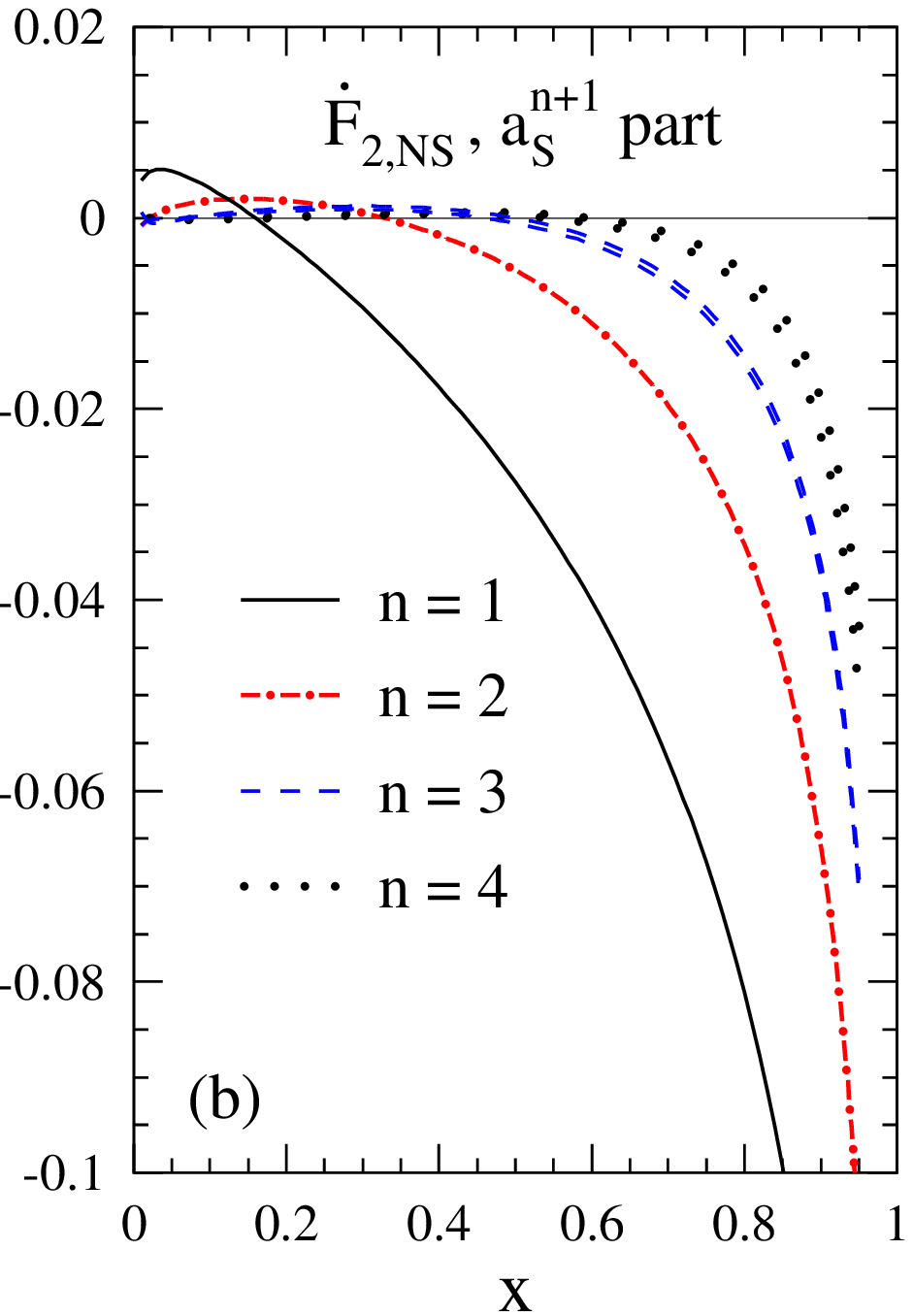,width=5.2cm,angle=0}
\epsfig{file=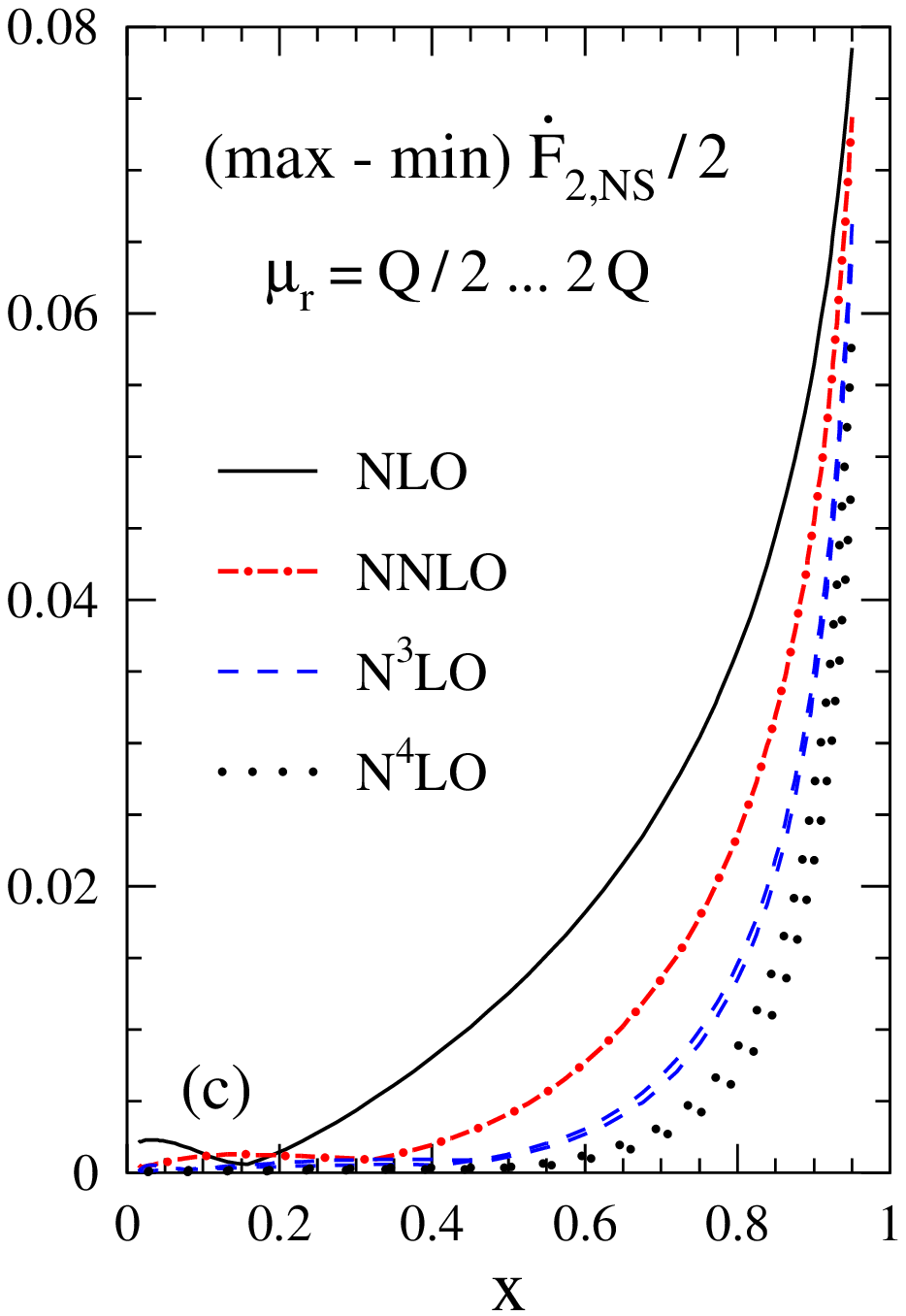,width=5.2cm,angle=0}}
\caption{
 {\bf (a)} The perturbative expansion of the logarithmic $Q^2$ 
 derivative of $F_{2,\rm NS}$ for the input (\ref{eq5}) for $N_f = 4$ 
 and $\mu_r^2 = Q^2$. 
 {\bf (b)} The N$^n$LO contributions, $n = 1\,\ldots\, 4$, to the 
 results shown in (a), and 
 {\bf (c)} their renormalization scale uncertainties. 
 }
\label{avf4}
\end{figure}

\newpage
The effect of the higher-order terms (using the Mellin inverse of 
Fig.~5(c) at N$^4$LO, see Fig.~12 of ref.~\cite{NV4} for the Pad\'e 
estimates of the corrections beyond N$^4$LO) is exemplified in 
Fig.~6 for the logarithmic $Q^2$ derivative of 
\beq
\label{eq5}
  F_{2,\rm NS}^{}(x, Q_0^2\,\approx\, 30 \mbox{ GeV}^2) \, =\, 
  x^{0.5} (1-x)^3 \:\: , \quad \alpha_s (Q_0^2)\, =\, 0.2 \:\: .
\eeq
Also shown are the renormalization-scale uncertainties estimated using 
the conventional interval $\frac{1}{4}\, Q^2 \leq \mu_r^2\leq 4\, Q^2$. 
The $\alpha_s^4$ (N$^3$LO) corrections and the N$^3$LO scale dependence 
are very small at $x < 0.6\,$; they both reach about 2\% and 5\% of the 
total results only at $x=0.65$ and 0.85, respectively. The corresponding
numbers at N$^4$LO read 1\% and~3\%.

Sample fits of $\alpha_s$ to the $Q^2$ derivatives at $0.05 \leq x \leq 
0.75$ for $Q^2 = Q_0^2$ (using the N$^3$LO predictions of 
Eq.~(\ref{eq5}) as model data) yield~\cite{NV4}
\bea
\label{eq6}
 \alpha_s(Q_0^2)^{\,}_{\rm NLO} \:\, & = &
 \: 0.2080 \,
 {\begin{array}{l} \scriptstyle + \:\: 0.021\\[-1.5mm]
 \scriptstyle - \:\: 0.013\:\:\end{array}} \: ,
 \quad\alpha_s(Q_0^2)^{\,}_{\rm NNLO} \, = \, 0.2010 \,
 {\begin{array}{l} \scriptstyle + \:\: 0.008\\[-1.5mm]
 \scriptstyle  - \:\: 0.0025\end{array}} \: ,
 \nonumber\\
 \alpha_s(Q_0^2)^{\,}_{\rm N^3LO}\:
 & = &
 \: 0.2000 \,
 {\begin{array}{l} \scriptstyle + \:\: 0.003 \\[-1.5mm]
 \scriptstyle  - \:\: 0.001\:\:\end{array}} \: ,
 \quad\alpha_s(Q_0^2)^{\,}_{\rm N^4LO} \:\, = \, 0.2000 \,
 {\begin{array}{l} \scriptstyle + \:\: 0.0015 \\[-1.5mm]
 \scriptstyle  - \:\: 0.0005 \end{array}} \: ,
\eea
where the errors include the above $\mu_r$ variation and, at N$^3$LO 
and N$^4$LO, the small approximation uncertainties. Corresponding 
results for $F_3$ can be found in ref.~\cite{NV4}. In both cases the 
N$^3$LO and N$^4$LO corrections, unlike the NNLO terms, do not cause
significant shifts of the central values, but just lead to a reduction
of the $\mu_r$ uncertainties which reach about $\pm 1\%$ at N$^3$LO.

{\sc Fortran} subroutines of our approximations of splitting functions
\cite{NV3} and non-singlet coefficient functions~\cite{NV4} at three 
loops, and of the parametrizations~\cite{NV12} of the two-loop 
coefficient functions~\cite{c2DIS} and the convolutions entering the 
physical evolution kernels~\cite{NV4} can be found at 
{\tt http://www.lorentz.leidenuniv.nl/$\sim\,$avogt}.

\section*{Acknowledgment}
\vspace*{-1mm}

This work has been supported by the European Community TMR research 
network `QCD and particle structure' under contract 
No.~FMRX--CT98--0194. 

\end{document}